\documentstyle{cupconf}

\ifoldfss
\else
  \ifnfssone
    \newmathalphabet{\mathit}
      \addtoversion{normal}{\mathit}{cmr}{m}{it}
      \addtoversion{bold}{\mathit}{cmr}{bx}{it}
    \newmathalphabet{\mathcal}
      \addtoversion{normal}{\mathcal}{cmsy}{m}{n}
    \else
    \ifnfsstwo
    \fi
  \fi
\fi

\def\Halpha{H$\alpha$}
\def\msun{M$_{\odot}$}
\def\etal{et al.}

\def\reff{$R_{\rm eff}$}
\def\sn{S$_N$}

\def\hexnumber#1{\ifcase#1 0\or1\or2\or3\or4\or5\or6\or7\or8\or9\or
 A\or B\or C\or D\or E\or F\fi }

\makeatletter
\ifx\CUP@mtlplain@loaded\undefined
\else
\fi
\makeatother
 \makeatletter
 \ifx\CUP@mtlplain@loaded\undefined
   \font\tenbmi=cmmib10 at 10pt
   \font\sevenbmi=cmmib10 at 7pt
   \font\fivebmi=cmmib10 at 5pt

   \newfam\bmifam
   \textfont\bmifam=\tenbmi
   \scriptfont\bmifam=\sevenbmi
   \scriptscriptfont\bmifam=\fivebmi
   
 \fi
 \makeatother
\ifnfsstwo

\fi
\ifnfssone

\fi
\ifoldfss

\fi

\mathchardef\varLambda="0103

\makeatletter
\ifx\CUP@mtlplain@loaded\undefined
\else
\fi
\makeatother
\makeatletter
\ifx\CUP@mtlplain@loaded\undefined
  \font\tenbms=cmbsy10
  \font\sevenbms=cmbsy10 at 7pt
  \font\fivebms=cmbsy10 at 5pt
  \newfam\bmsfam
  \textfont\bmsfam=\tenbms
  \scriptfont\bmsfam=\sevenbms
  \scriptscriptfont\bmsfam=\fivebms

  \edef\bsy@{\hexnumber\bmsfam}
  \mathchardef\bnabla="0\bsy@72
\fi
\makeatother
\def\etal{\mbox{\it et al.}}

\title[The Formation of Star Clusters]{The Formation of Star Clusters}

\author[B. C. Whitmore]%
{B\ls R\ls A\ls D\ls L\ls E\ls Y\ns C.\ns W\ls 
H\ls I\ls T\ls M\ls O\ls R\ls E$^1$}

\affiliation{$^1$Space Telescope Science Institute, 3700 San Martin Drive, Baltimore, MD, 21218}

\begin{document}
\ifnfssone
\else
  \ifnfsstwo
  \else
    \ifoldfss
      \let\mathcal\cal
      \let\mathrm\rm
      \let\mathsf\sf
    \fi
  \fi
\fi

\maketitle

\begin{abstract}
The ability of HST to resolve objects ten times smaller than possible
from the ground has rejuvenated the study of young star clusters.  A
recurrent morphological theme found in nearby resolved sytems is the
observation of young (typically 1 - 10 Myr), massive (10$^3$ - 10$^4$
\msun\/), compact ($\rho$ $\approx$ 10$^5$ \msun\/ pc$^{-3}$) clusters
which have evacuated the gas and dust from a spherical region around themselves. New stars
are being triggered into formation along the edges of the envelopes,
with pillars (similar to the Eagle Nebula) of molecular gas streaming
away from the the regions of star formation. The prototype for these
objects is 30 Doradus (Figures 1 \& 2). Another major theme has
been the discovery of large numbers of young (typically 1 - 500 Myr),
massive (10$^3$ - 10$^7$ \msun\/), compact star clusters in merging,
starbursting, and even some barred and spiral galaxies.  The brightest
of these clusters have all the attributes expected of protoglobular
clusters, hence allowing us to study the formation of globular
clusters in the local universe rather than trying to ascertain how
they formed $\approx$ 14 Gyr ago. The prototype is the Antennae Galaxy
(Figures 3 \& 4).

\end{abstract}

\firstsection % if your document starts with a section,
              % remove some space above using this command.
\section{Introduction}

\subsection{Hubbles' first six months}

The discovery of spherical aberration in the summer of 1990 raised
serious questions about the ability of HST to do the unique science it
was designed for, and caused general consternation throughout the
astronomical community. Early HST observations 
of compact star clusters in 30 Doradus and NGC 1275 played  pivotal
roles in demonstrating that HST, even in its crippled state, could
produce stunning results that were impossible to obtain from the
ground. This provided a much needed shot in the arm for the
Hubble project.  In addition, image deconvolution techniques developed to
reconstruct the 30~Doradus images showed that most of the compromised
resolution could be restored for bright objects.

Before the HST image of 30 Doradus came out in the summer of 1990,
several papers had argued that R136 (the central object in the 30
Doradus cluster) was a single star with a mass of several thousand
solar masses (e.g., Cassinelli, Mathis \& Savage 1981). By the time HST was launched, speckle observations by
Weigert  \& Baier (1985) had resolved the central region into roughly a dozen
individual stars. However, concerns about possible artifacts
introduced by the speckle technique, and limitations imposed by the
small field of view, limited the impact of the results. As the saying
goes, ``a picture is worth a thousand words'', and it was the
spectacular direct images obtained with the WFPC1 on HST that first
made it clear just how rich this cluster really was. For example, Hunter \etal\/ (1995)
identified over 3500 stars in the central $\approx$ 8$^{''}$ alone. 

HST was built to make new discoveries, and one of the first was the
discovery of young compact clusters in NGC 1275, the central galaxy in
the Perseus cluster (Holtzman \etal\/ 1992). This demonstrated
that HST had no peers when it came to
detecting compact, point-like objects against a bright
background. Objects that were impossible to see from the ground
suddenly became visible. Holtzman \etal\/ (1992) also made the rather daring
assertion that the young clusters were {\it protoglobular} clusters
formed by a merger event. This was the catalyst for what has become a
major cottage industry for Hubble, and provides one of the primary
topics for this review.

\begin{figure*}
\includegraphics{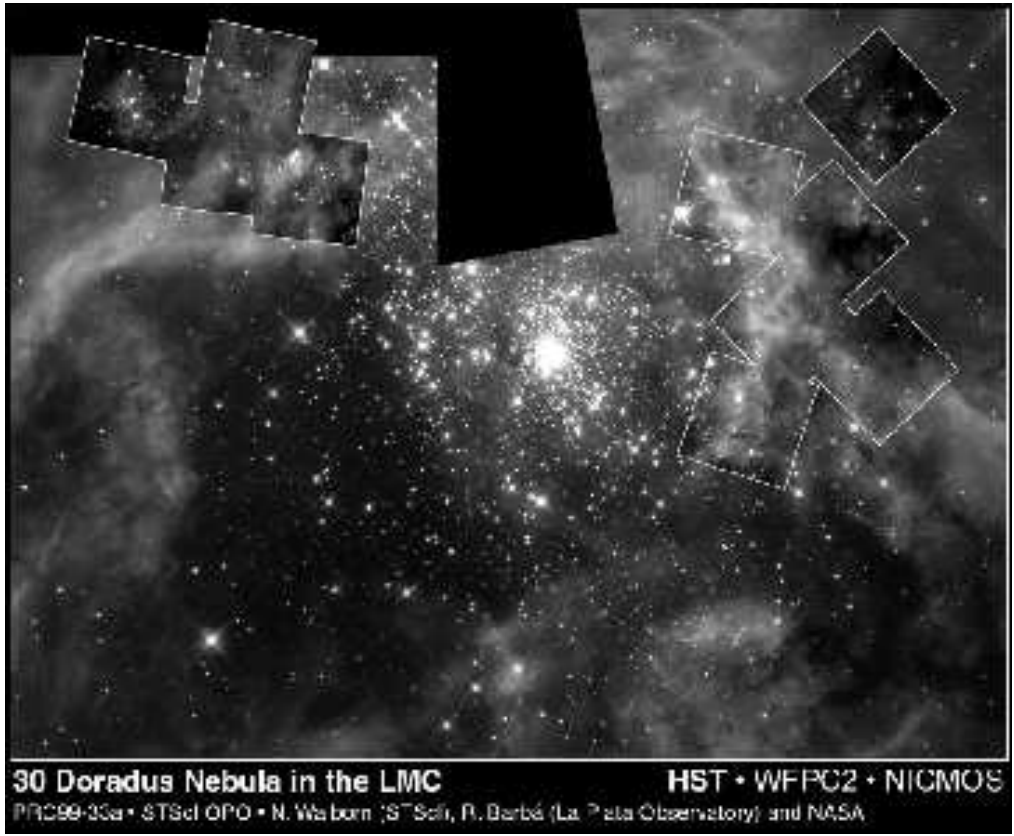}
\vspace{12.5cm}
\caption{Combined WFPC2 (background) and NICMOS (square inserts) image
of 30 Doradus from a 1999 press release by Walborn and Barba. 
The central star cluster is R136. }
\end{figure*}

\subsection{Motivation}

There are three basic reasons why the HST observations of young 
star clusters caught the attention of the astronomical community.

1) They provide insight into the mechanisms of star formation, the
most fundamental process in astronomy.  While we have very detailed
models of the structure and evolution of stars (e.g., isocrones in the
HR diagram), we have only sketchy ideas of how stars form to begin
with.  Some of the basic questions that remain to be solved are: Is
the initial stellar mass function the same in all environments?  Can
star formation trigger subsequent star formation?  Are all stars
formed in groups and clusters?

An obvious approach to solving these questions is to go to where 
lots of stars are forming, such as merging and starbursting
galaxies. When we do this we find that a large fraction of the star
formation is in the form of massive, compact star clusters.  
Understanding how these clusters form should go a long ways toward
understanding star formation in general.

2) They allow us to study the formation and evolution of globular star
clusters in the local universe rather than trying to ascertain how
they formed $\approx$ 14 billion years ago. An analogy is often made
between the study of old globular clusters and the study of fossils on
the earth, since both provide an evolutionary record. Given
the opportunity, I believe many paleontologists would switch fields if
they could go to where the dinosaurs are still living.

3) They are relevant to the question of whether spiral galaxies can
merge to form elliptical galaxies. One of the primary objections to
this hypothesis was raised by van den Bergh (1990), who pointed out
that spirals have fewer globular clusters per unit luminosity than
ellipticals. It now appears that globular clusters can be formed by
mergers of gas rich systems (\S 3.1), providing a natural
explanation for this difference.

\subsection{Nomenclature}

For historical reasons, a wide variety of names are currently being
used to describe what are physically similar objects. Researchers
studying merging galaxies often use names such as ``protoglobular
cluster'' or ``young globular clusters''. This is because the focus
has been on the question of whether merging galaxies can form globular
clusters, in response to van den Bergh's (1990) criticism of the
merger hypothesis for the origin of elliptical galaxies.  On the other
hand, researchers studying nearby starburst galaxies generally use the
term ``super star clusters'', a term that was first introduced by Arp
\& Sandage (1985) when referring to the dominant cluster in the starburst dwarf
galaxy NGC 1569.  Other names in use include ``blue populous
clusters'' (in the LMC) and ``young massive clusters'' (in normal spirals;
e.g., Larsen \& Richtler 1999).

None of these names really ring true, however, which is why one has
not become the standard. While there is good evidence that some of the
brightest compact clusters become globular clusters, it is obvious
that they do not all become globular clusters, since the specific
globular cluster frequency in merger remnants would then be too high.
The main objection to the term ``super star cluster" is that while
they may be very luminous for a short period of time, their masses,
which are a more fundamental property, are not ``super''. They are
instead similar to normal globular clusters or open clusters.  In
essence, there are probably no major physical differences between
these various clusters; we are simply seeing young globular clusters,
open clusters, and associations at different stages of their evolution
or in different environments than we are use to seeing them in the
Milky Way.

The defining properties for most of the objects discussed in this review are
that they are young, compact, and the brightest are ultra luminous in
comparison to old globular clusters. I will simply call them young
compact star clusters.

\subsection{Goals}

I have three goals for this review. The first is to highlight the
primary contributions that HST has made to the study of the formation
of compact star clusters (\S 2 \& 3).  There will be an obvious bias
towards observations of young compact clusters in merging
galaxies, since that is what I know the most about, but I will also
highlight some of the major results for nearby clusters such as
30 Doradus. The reader is referred to reviews in this volume by Harris (old globular
clusters), Bally (star formation), and 
Leitherer (starbursts) for related discussions. 
The second goal is to provide a compilation of
the literature  on young unresolved compact clusters, as discussed in \S 4. The third goal
is to use the compilation to examine some of the demographics of
young cluster formation (\S 4).

\section{Nearby Resolved Star Clusters}

The techniques for studying star clusters that can be resolved into
individual stars are much different than the techniques available for
studying more distant clusters. For example, we can determine the
stellar luminosity and mass functions, the color-magnitude diagram for
the stars, and the radial density profile. For the more distant
clusters some of the typical tools are the specific globular cluster frequency,
the cluster luminosity function, and color-color diagrams. Before HST,
the dividing line between these two regimes was essentially the edge
of the Milky Way galaxy.  One of the promises of HST was that it would
push this dividing line roughly ten times farther away, allowing us to
study the clusters in the Magellanic Clouds, and to a lesser extent
the nearby galaxies in the Local Group, with the same techniques we
have been using for the clusters in the Milky Way. This has indeed
been the case. In this section we describe three of the primary
HST results for the nearby resolved clusters.
Table 1 lists the number, luminosity of the brightest cluster, size, mass, age, and 
power law index for the luminosity function for a variety of young star clusters that are discussed in this review.

% \begin{table}[htb]
\begin{table}
\begin{center}
\caption{Properties of Young Compact Star Clusters (approximately in order of distance)}
% \begin{tabular}{lll}
\begin{tabular}{p{1.0in}p{0.5in}p{0.5in}p{0.5in}p{0.8in}p{0.8in}p{0.5in}}
\hline 
 galaxy  & N & M$_V$(bright) & \reff & mass\ & age & $\alpha$\\
\hline
near Gal. Center & 2 & --- & --- & 10$^4$ \msun\/ & 3 Myr & --- \\
LMC  &  8 & -11.3 & 2.6 pc & --- & 3 Myr & ---\\

M82 & 100 & -14.5 &  3.5 pc & --- & 100 Myr & ---\\

HE 2-10 & 76 & -12.7   &  3 pc & 10$^3$ - 10$^5$ \msun\/  & --- & -1.7 \\

ESO 338-IG04 & 112 & -15.5 & --- &  10$^3$ - 10$^7$ \msun\/ & 10-10,000 Myr    & --- \\ 

% NGC 253 & 4 & -15 & --- & 1.5 $\times$ 10$^6$ \msun\/ & & --- \\) 

NGC 1569 & 7 &  -13.9  & 2.2 pc & --- & 15 Myr & ---\\

NGC 5253 & 6 & -11.1  & --- & 10$^6$ \msun\/& 2.5 Myr & ---  \\

NGC 1705 & 36 & -13.7  & 3.4 pc & ---  &  15 Myr & ---\\

NGC 1741 & 314 & -15  & --- & 10$^4$ - 10$^6$ \msun\/ & 4 Myr & -1.85 \\

ESO 565-11 & 700 & -13.4 & --- & --- & 4 - 6 Myr & -2.2\\

NGC 1097 & 88	& -13.8 & 2.5 pc & --- & --- & --- \\

NGC 4038/39 & 800 & -15.8  & 4 pc &  10$^3$ - 10$^7$ \msun\/ & 1 - 500 Myr & -2.1 \\

NGC 3256 & 1000 &-15 & 5 - 10 pc  & --- & --- & -1.8 \\

NGC 3256 tail  &	50 & -10  & --- & --- & ---  & ---\\

NGC 3597 & 700 & -13.2 & --- & --- &  --- & -2.0\\

NGC 7252 & 500 & -16.2  & 5 pc & 10$^4$ - 10$^8$ \msun\/ &  600 Myr & -1.8\\

NGC 1275 & 800 & -15.8  & --- & 10$^4$ - 10$^8$ \msun\/ & 0.1 - 1.0 Gyr & -1.9 \\

NGC 3921 & 102 & -14  & $<$ 5 pc & --- & 500 Myr  & -2.1 \\

\end{tabular}
\end{center}

Notes to Table 1: See Table 2 for references. N is the number of clusters (to 
M$_V$ $\approx$ -9 mag in most cases).  M$_V$(bright) is the M$_V$ magnitude for the brightest cluster. \reff\/ is the average effective radius for the clusters. $\alpha$ is the power-law index for the luminosity function.
\end{table}

\subsection{Young Clusters near the Center of the Milky Way}

The Near-infrared Camera and Multi-object Spectrometer (NICMOS) was
used by Figer \etal\/ (1999) to penetrate the dust toward the center
of the Milky Way in order to study two remarkable young clusters near the
Galactic Center. Based on turnoffs in the color-magnitude diagrams
they estimate that the Arches cluster is 2 $\pm$ 1 Myr old and the
Quintuplet cluster is 4 $\pm$ 1 Myr old.  Based on number counts and
an extrapolation to 1 \msun\/, they estimate the masses of the clusters
are $\approx$ 10$^4$ \msun\/, and the densities are $\approx$ 10$^5$ \msun\/ pc$^{-3}$. 

The existence of these clusters was somewhat unexpected; one would think that
the strong tidal shear produced by the central black hole, the high
velocity dispersion, and the strong magnetic field would make this a
hostile environment for forming young massive clusters. However, given
that starburst galaxies are able to support prodigious rates of star
formation near their centers, perhaps we should not be surprised that
star clusters can also form near the Galactic Center.

Portegies Zwart (2000) performed n-body simulations which indicate that the
two clusters should dissolve after 10 - 60 million years in the tidal
field of the Galaxy. They also point out that the stellar density near
the center of the Galaxy is so high that the clusters would only be
distinguishable for a short fraction of their existence, as low as 5
\% in some models. Based on this result they conclude that the
Galactic Center may be hiding between 10 and 40 clusters which are
similar to the Arches and Quintuplet clusters, but slightly older and
less compact. Similar simulations by Kim
\etal\/ (1999) suggest that very massive stars play an important
role in the evolution of these clusters because relaxation  and mass
segregation times are comparable to or even smaller than the lifetimes
of the stars.

These clusters are reminiscent of the young clusters found in the
central 1.5 kpc of the merger NGC 7252. Miller \etal\/ (1997) found
$\approx$ 40 such clusters, all less than $\approx$ 10 Myr years old based
on the (U-B) vs. (V-I) diagram. Hence, it looks like the centers of
galaxies may actually be good places to make star clusters, but few if
any will survive very long. This suggests that a sizeable fraction of
the field star population may have been formed in clusters. 

\subsection{The Initial Stellar Mass Function (IMF)}

The light from young star clusters is dominated by ultraluminous O and
B stars. However, if these were the only stars the cluster would not
be stable since the massive stars are destined to go supernova,
returning their mass to the interstellar medium.  Hence, the identification of low
mass stars is critical for the survival of the compact star
clusters. In addition, the question of whether the IMF is universal,
or is instead a function of the environment, is a question which is
currently being hotly discussed (see reviews by Scalo 1998, Larson 1999,
and Elmegreen 1999). I will briefly comment on a few examples where
HST observations of young clusters are relevant to this debate.

The initial stellar mass function is generally parameterized as a power law with an index $\Gamma$. The cannonical value of $\Gamma$ for field
stars in the Milky Way is -1.35 (Salpeter 1955). There is a lively
debate over the question of how universal $\Gamma$ is, and in
particular, whether $\Gamma$ is the same in clusters and in the
field. Several authors have argued that the formation of high mass
stars may be favored in starburst regions, which would result in a
lower value of $\Gamma$.

Initial measurements of the IMF in 30 Doradus were made by Malumuth \&
Heap (1994; found over 800 stars in the inner 8$''$ using the WFPC1),
and Hunter \etal\/ (1995; found over 3500 stars in the inner
8$''$ using WFPC2).  More recently, Massey and Hunter (1998) measured
the IMF down to 2.8 M$_{\odot}$, and find that despite the largest
number of high mass and luminosity stars ever seen, the IMF is
completely normal. This implies that the IMF is the same over several
orders of magnitudes in density, from field stars to starburst
clusters like 30 Doradus.  The large number of O stars is simply a
result of the youth ($\approx$ 2 Myr) and richness of the cluster.  A
more recent study of WFPC2 images by Sirianni \etal\/ (2000) detects
stars in 30 Doradus roughly 1 magnitude deeper than Massey and Hunter,
and argues that these are pre-main sequence stars in the mass range 0.6
- 3 \msun\/. They construct the IMF in the range 1.35 - 6.5 \msun\/ and
find a flattening below $\approx$ 2 \msun\/. While there are several
examples of a flat IMF for stars less massive than 1 \msun\/ (see
Larson 1999), this is the first example of a flattening above this
point.

Another possible example of a relatively flat IMF, but this time at
the high mass end, are the results of Figer \etal\/ (1999) for the
Arches cluster, one of the two massive young clusters near the
Galactic Center.  They find $\Gamma$ = -0.7 when fitting over the
range 6 - 125 \msun\/ for stars in an annulus from 3$''$ to
7.5$''$. However, incompletion caused by crowding make this a
difficult measurement at the faint end of the mass function.  For
example, Figer \etal\/ note incompletion fractions of 50 \% for stars
up to 35 \msun\/ in the inner 3$''$. In addition, they find a very
flat value of $\Gamma$ = -0.2 in the region 2.5$''$ to 4.5$''$, which
may be due to truncation of the faint end by crowding. If we stick to
the brighter end of the mass function from 16 to 125 \msun, and only
use stars in an annulus from 4.5$''$ to 7.5$''$, $\Gamma$ $\approx$
-1.1 , still low but not too different from the Salpeter IMF.

\begin{figure*}
\includegraphics{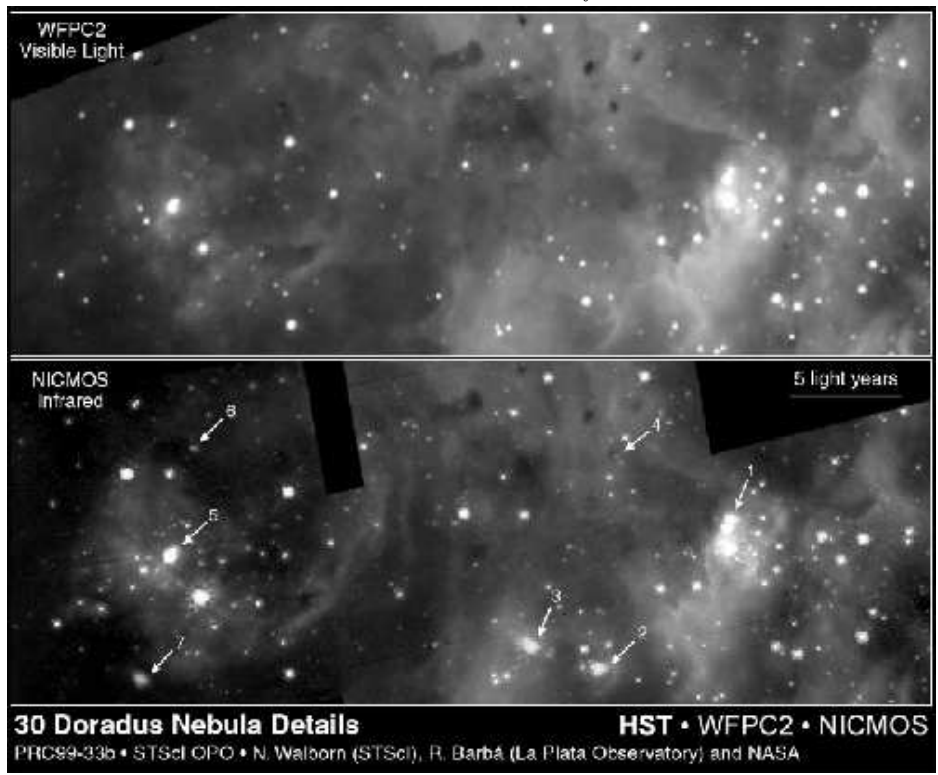}
\vspace{10.2cm}
\caption{WFPC2 and NICMOS images in an outlying region of 30 Doradus (i.e., the
upper left square inserts in Figure 1) from a 1999 press release by Walborn and Barba. Note the young stars still embedded in dust which are only visible on the NICMOS image (e.g., objects identified as 2, 3, 6 and 7). Also notice the pillar-like dust feature around object 1 pointing away from the central star cluster R136 (see Figure 1 for context).}
\end{figure*}

Hence, while there is tentative evidence for deviations in the IMF for
some environments, the near uniformity over an extremely wide range of
environment is quite remarkable. Larson (1999) concludes ``this large
body of direct evidence does not yet demonstrate convincingly any
variability of the IMF, although the uncertainties are still
large. Some indirect evidence based on the photometric properties of
more distant and exotic systems suggests that the IMF may vary in
extreme circumstances, possibly being top-heavy in starbursts and high
redshift galaxies.'' We also note that low-mass stars are clearly
forming in R136 (e.g., down to 0.6 \msun\/; Sirianni \etal\/ 2000),
and in other young nearby compact clusters such as NGC 3603 (i.e., no
evidence for a flattening down to 1 \msun\/, Eisenhauer \etal\/ 1998, ). Hence
concerns that the young clusters will be unstable due to the absence
of low mass stars appear to be unfounded.

\subsection{Triggered Star Formation }

30 Doradus has been called a ``Starburst Rosetta'', since it is the
nearest example of a young massive starburst cluster and hence can be
studied in unique ways that are not possible for its more distant
counterparts.  Until
recently, 30 Doradus was believed to be a well evolved H II region,
with no current star formation going on.  Early indications that this
was not true were the discovery of an H$_{2}$O maser (Whiteoak 1983)
and four luminous IR protostars (Hyland \etal\/ 1992).  More recently,
Walborn \& Blades (1997) used optical spectral classification of 106
OB stars to show the presence of five distinct stellar populations in
30 Doradus, with ages ranging from $\approx$ 1 to $\approx$ 10
Myr.  Hence, the simple models of either continuous or single-burst
star formation appear to be incorrect. It now appears that a starburst
in one area can trigger the formation of stars in a nearby region.

Corroborating evidence for this picture has been obtained by Walborn
\etal\/ (1999b) using NICMOS, as shown in Figure 1. A roughly
spherical shell of ``second generation'' star formation, including a
host of newly discovered IR sources, can be seen around the central
R136 concentration (Figure 2). Several massive dust pillars are found streaming
away from R136, similar to the famous HST image of the Eagle Nebula.
At the heads of these pillars are the sites where active star formation
is currently being triggered. The molecular gas revealed by the dust provides
the raw material for the star formation. Scowen \etal\/ (1998) shows
that these pillars are indeed very similar to the pillars in the Eagle
Nebula. This same picture of a bright compact central cluster which has
evacuated a roughly spherical nebular envelope around it can be seen
in a number of the HST press releases (e.g., NGC 604, N11 in the LMC,
NGC 4214).

\section{Distant Unresolved Star Clusters}

In this section we move further out to where it becomes difficult to
resolve the individual stars, except in extraordinarily extended
clusters such as knot S in NGC 4038/4039 (the ``Antennae Galaxies'',
Whitmore \etal\/ 1999). Historically, most of the early HST
observations of compact star clusters were in merging galaxies,
followed shortly by similar observations in starburst galaxies. More
recently, young compact clusters have also been found in other environments,
including barred galaxies, tidal tails, and normal spiral galaxies.

\subsection{Young Compact Star Clusters in Merging Galaxies}

The primary question for many of the early HST studies of merging
galaxies was whether globular clusters were being formed.  This
possibility was proposed by Schweizer (1987) and Burstein (1987),
primarily to address van den Bergh's (1990) objection to the merger
model based on the higher specific frequency of globular clusters in
elliptical galaxies. Ashman and Zepf (1992) and Zepf and Ashman (1993)
further developed these ideas, and made predictions about the
bimodality of the metallicity histogram of globular clusters that
should result if most ellipticals are formed by merging spirals.

A hint that young globular clusters might be formed in mergers was
provided by Schweizer's (1982) observations of six unresolved bluish
knots in the merger remnant NGC 7252. However, with so few objects he
could not be sure they were not simply field stars. Lutz (1991)
observed roughly a dozen blue, point-like objects in the merger remnant
NGC 3597, but was not able to resolve the objects and hence could not
be certain they were not associations or giant HII regions.

As briefly discussed in \S 1, the HST observations of NGC 1275 by
Holtzman \etal\/ (1992) provided the original breakthrough and was the
primary catalyst in this field.  They discovered a population of about
60 blue compact clusters, and suggested that they were protoglobular
clusters which formed $\leq$ 300 Myr years ago during a merger of NGC
1275 with another galaxy.  Unfortunately, NGC 1275, the central
cooling-flow galaxy in the Perseus cluster, is such a peculiar galaxy
that it is not clear which of its peculiarities is responsible for the
formation of the young clusters (e.g., see Richer \etal\/ 1993, who
suggested that the cooling flows are responsible for the formation of
the clusters).

\begin{figure*}
\includegraphics{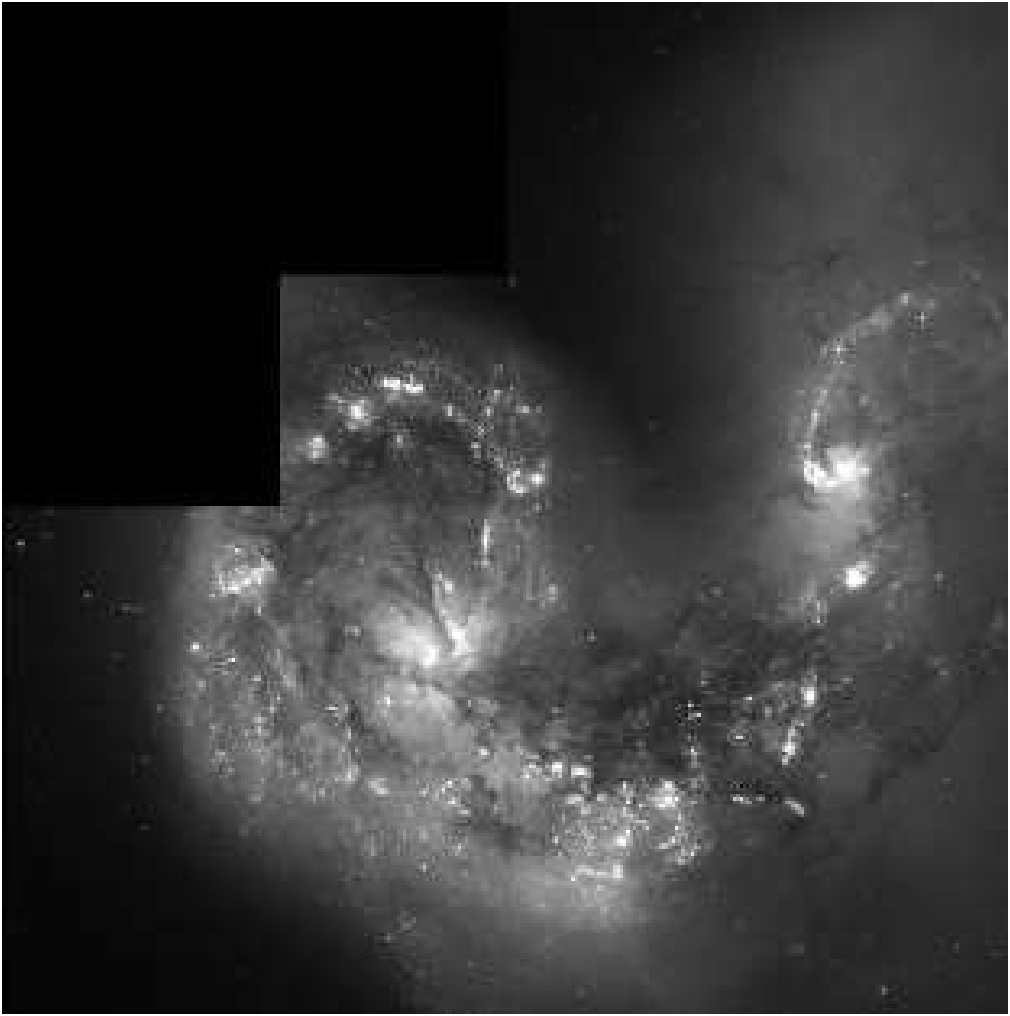}
\vspace{14.2cm}
\caption{Image of the Antennae Galaxies (NGC 4038/4039) from Whitmore \etal\/ (1999).}
\end{figure*}

Whitmore \etal\/ (1993), using WFPC1 observations  of the prototypical merger remnant NGC 7252 (Toomre 1977), found a
population of about 40 blue point-like objects with luminosities and
colors nearly identical to those found in NGC 1275. Unlike NGC 1275,
with all its peculiarities, NGC 7252 is an isolated galaxy which
therefore provided a much cleaner connection between the formation of
young star clusters and merging galaxies. Whitmore \& Schweizer (1995)
followed this up with pre-refurbishment observations of another
prototypical merger, NGC 4038/4039 (see
Figures 3 \& 4).  Over 700 young star clusters were found in this
galaxy. Subsequent observations of both these galaxies using WFPC2
(NGC 7252 - Miller \etal\/ 1997; NGC 4038/4039 - Whitmore \etal\/ 1999)
have increased the numbers of cluster candidates tenfold.

Roughly 30 different gas-rich mergers have now been observed with
HST, as summarized in Table 2. In all cases young massive compact
clusters have been observed, {\it the brightest of which have the
luminosities, colors, sizes, masses, distributions and spectra that we
would expect for globular clusters with ages in the range 1 to 500
Myr}. A few of the key observations are described below, but the reader
is referred to the papers listed in Table 2 for the
details.

\begin{figure*}
\includegraphics{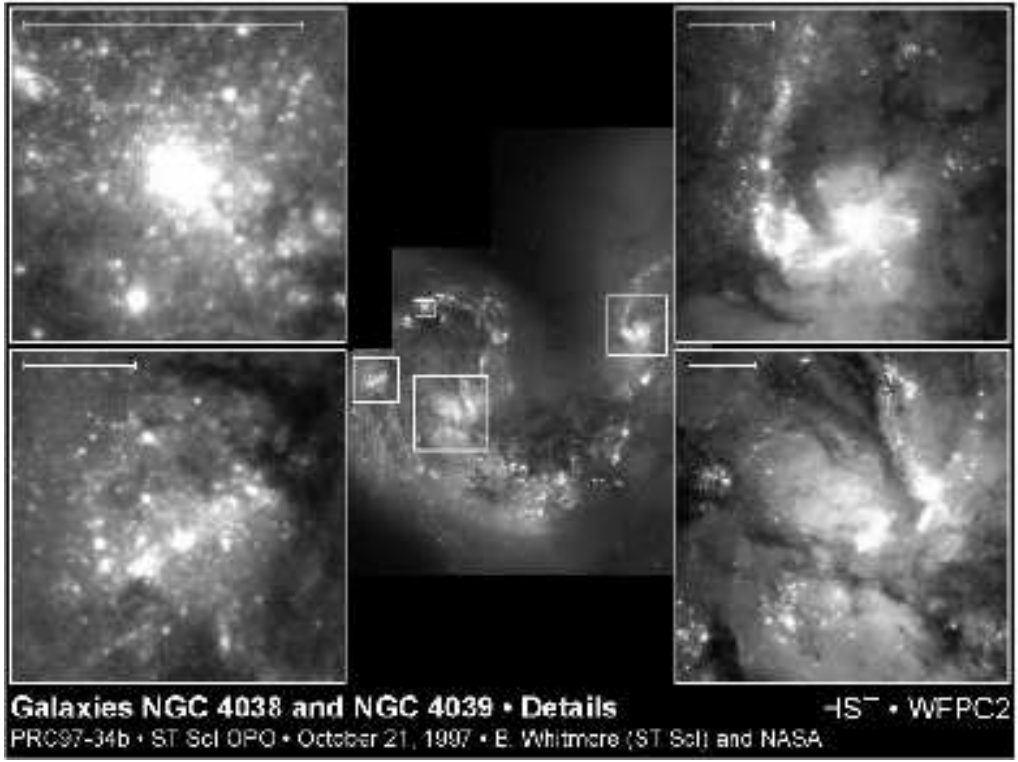}
\vspace{11.5cm}
\caption{Blowup of two of the brightest clusters in the Antennae (left) 
and the central regions of the two galaxies (right) 
from Whitmore \etal\/ (1999).}
\end{figure*}

\subsubsection{Luminosities and Colors}

The luminosities of young globular clusters with ages $\approx$ 10 Myr
should be $\approx$ 5 - 6 magnitudes brighter than classical old
globular clusters, according to the Bruzual and Charlot (1996)
models. The models also predict that young globular clusters should be
$\approx$ 1.0 - 1.2 magnitude bluer in $(V-I)$. Figure 5 shows that
this is indeed the case. It also shows how the luminosities and colors
of the clusters can be used to age date the clusters. NGC 4038/4039 is
clearly the youngest merger remnant, with the mean age of the clusters
$\approx$ 30 Myr and many clusters only a few Myr old. The clusters in
NGC 3921 and
NGC 7252 are roughly 500 Myr old while NGC 3610 appears to be a 4
$\pm$ 2 Gyr merger remnant (Whitmore \etal\/ 1997) which may provide
the missing link between young mergers and old ellipticals. The case is less
certain for NGC 1700, although Brown \etal\/ (2000) have recently
claimed that this galaxy also has a population of metal-rich clusters
that are 3 $\pm$ 1 Gyr old. Goudfrooij \etal\/ (2000) also found a
population of $\approx$ 3 Gyr clusters in NGC 1316.

\begin{figure*}
% \special{psfile=bcw4.ps hoffset=440 voffset=160 hscale=80 vscale=80
% angle=180}
\includegraphics{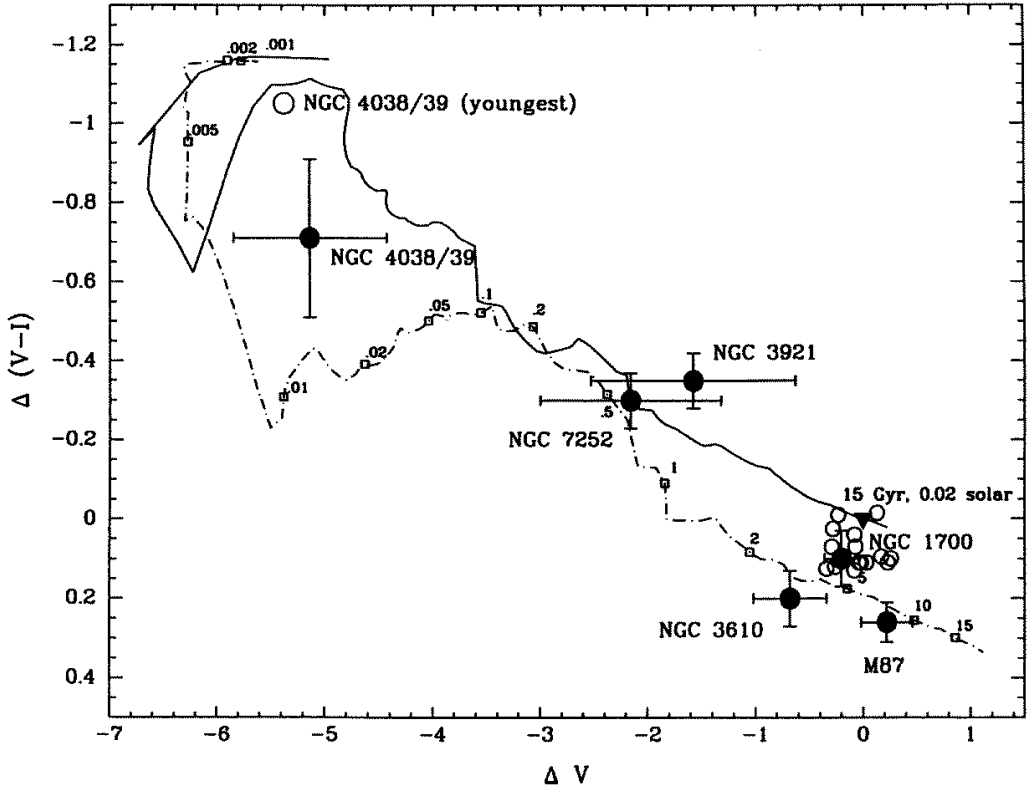}
\vspace{11.0cm}
\caption{Plot of the evolution in luminosity ($\Delta$ V) and in color 
($\Delta$ (V-I)) of star
clusters, based on the  Bruzual-Charlot (1996) tracks for a metal-poor
population (solid line) and a solar metallicity population (dashed-dot line).
The values are normalized to an old, metal-poor population (filled triangle).
Ages in Gyr for the solar metallicity track are marked with squares.
See the original article for further details (Whitmore \etal\/ 1997). 
}
\end{figure*}

\subsubsection{Sizes}

The ability to measure sizes using HST has been critical to the identification of the young clusters in mergers as candidate
globular clusters.  Ground based observations, such as those of Lutz
(1991), were inconclusive, since they were not able to resolve the
clusters to determine whether they were associations or HII regions,
with \reff\/ $\approx$ 100 pc, or compact clusters similar to the
globular clusters in the Milky Way, with \reff\/ $\approx$ 3 pc (van den
Bergh, 1996).  Early HST observations using WFPC1 indicated that the
clusters were compact, with \reff\/ $\approx$ 10 pc (Whitmore \&
Schweizer, 1995). However, van den Bergh (1995) argued that this was
too large, and the clusters were more likely to be open clusters.
Meurer \etal\/ (1995) found that the compact clusters he was studying in very
nearby starburst dwarfs were smaller, with \reff\/ $\approx$ 2 pc. He
suggested that the apparently larger values in the Antennae were due
to poorer resolution and crowding.

Recent observations using the WFPC2 (corrected for spherical
aberration) have removed this concern.  Several authors have recently
measured \reff\/ for young clusters in mergers in the range 3 - 6 pc
(i.e., NGC 3921 - Schweizer \etal\/ 1996, NGC 7252 - Miller \etal\/ 1997,
NGC 3610 - Whitmore \etal\/ 1997, and NGC 1275 - Carlson \etal\/
1998). Perhaps the best case is for the Antennae galaxies as measured
by Whitmore \etal\/ (1999), since this is the nearest of the
prototypical mergers and the observations were made with subpixel
dithering which improves the resolution still further. They find the
median effective radii for the clusters is \reff\/ = 4$\pm$1 pc,
similar to or slightly larger than those of globular clusters in the
Milky Way.

\subsubsection{Ages}

Ages for the clusters have been estimated in a variety of manners.
Figure 5 demonstrates how the luminosities and colors can be used to estimate
the ages, as already discussed in \S 3.1.1.  More precise age estimates
are possible with more colors, and provide an independent means of
solving for the age and the reddening caused by dust.  For example,
Whitmore \etal\/ (1999) use UBVI photometry and reddening-free Q
parameters to determine ages for the clusters in the Antennae (Figure 6). They
find evidence for four populations of clusters, ranging in age from
$<$~5 Myr to 500 Myr. They also isolate a population of old globular
clusters in this galaxy. {\it Hence, it appears that we can study the
entire evolution of globular clusters in this single galaxy.}  This is
consistent with the simulations  of Mihos, Bothun, \&
Richstone (1993) who find that the merger process takes several hundred million
years to complete,  hence producing clusters with a wide range of ages. 

\Halpha\/ can be used in two ways to estimate the ages of the younger
clusters. The existence of \Halpha\/ emission itself indicates that a
cluster is $<$ 10 Myr, since the O and B stars required to ionize the
gas only live for this long (e.g., see the Leitherer \& Heckman 1995
models). The second method is to use the size of the \Halpha\/ ring
around a cluster. Whitmore \etal\/ estimate that the clusters in the
western loop of NGC 4038 are 5 - 10 Myr, since many of them have rings
with diameters of $\approx$ 100 - 500 pc and measured expansion velocities $\approx$ 25 - 30 kms$^{-1}$ (Whitmore \etal\/ 1999). The clusters in the
overlap region appear to be $<$ 5 Myr old, since the rings are smaller or
non-existent in this region.

The most accurate method of estimating ages is to obtain spectra. Zepf
\etal\/ (1995) obtained spectra of the brightest cluster in NGC 1275
which showed strong Balmer absorption lines, typical of A stars. They
estimate ages of 500 Myr for the clusters, although ages from 100 -
900 Myr cannot be ruled out. Schweizer \& Seitzer (1998) obtained
UV-to-visual spectra of eight cluster candidates in NGC 7252. Six of the
clusters have ages in the range 400 - 600 Myr, roughly consistent with
the mean photometric age estimate of 650 Myr from Miller \etal\/
(1997). One cluster turned out to be an emission-line object with an
age estimate of $<$ 10 Myr, indicating that cluster formation is still
going on at a low level even in the outer parts of the
galaxy. Whitmore \etal\/ (1999) obtained GHRS spectra of two clusters
in the Antennae with age estimates of 3 $\pm$ 1 Myr and 7 $\pm$ 1 Myr,
in good agreement with the estimates based on the UBVI colors and the
\Halpha\/ morphology.

\begin{figure*}
\includegraphics{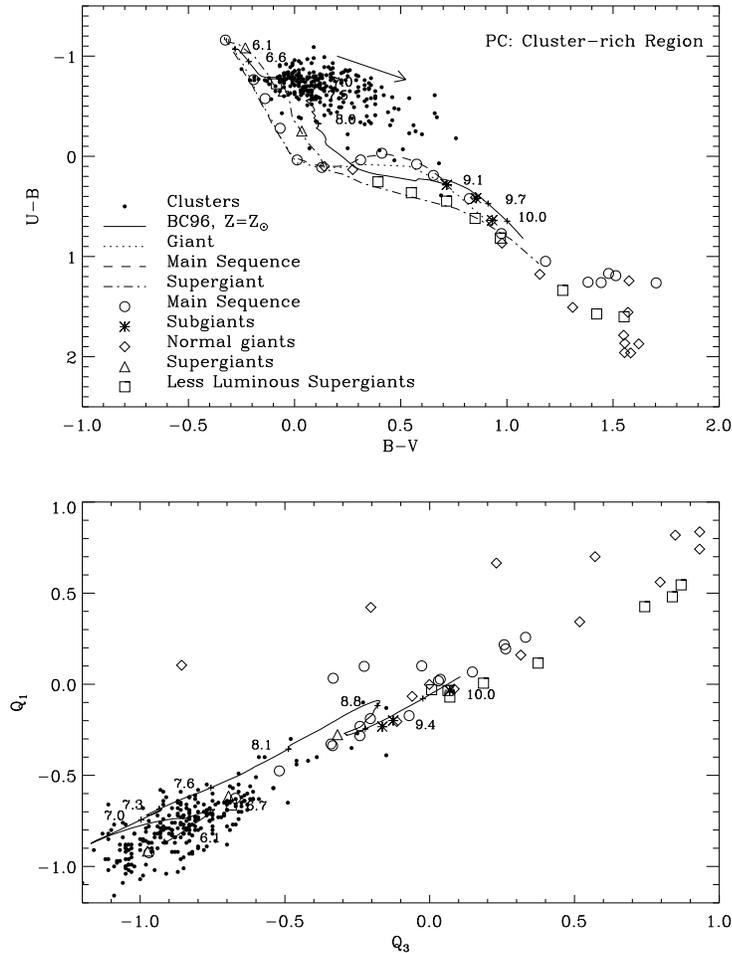}
\vspace{13cm}
\caption{Color-color diagram and reddening-free Q parameter diagram for clusters in the Antennae. The numbers  on the plots are the values of log(age). See Whitmore \etal\/ (1999) for details. }
\end{figure*}

The youngest clusters appear to be very red objects, which Whitmore \&
Schweizer (1995) suggested were only now emerging from their dust
cocoons. Several of these have recently been identified as strong IR
sources (Vigroux \etal\/ 1996, Mirabel \etal\/ 1998, Wilson \etal\/
2000, Gilbert \etal\/ 2000, Mengel \etal\/ 2000). In fact, the brightest IR source in the Antennae is one of
these very red objects (W80), rather than the nucleus of one of the
two galaxies.  Wilson \etal\/ (2000) find three separate molecular
clouds around W80 within a region of 1 kpc$^{2}$, and suggest that
cloud-cloud collisions may play an important role in cluster
formation.  However, the lack of similar morphologies for the other
very red objects suggest that this may not be the universal mechanism.

\subsubsection{Mass}

Mass estimates of young compact clusters have been made in two
ways. The first is based on the luminosity and color of the clusters
using stellar population models such as Bruzual \& Charlot (1996).
These estimates generally range from 10$^3$ to 10$^7$ \msun\/ (see
Tables 1 and 2), in good agreement with old globular clusters with a
mean of 2 $\times$ 10$^5$ \msun\/ (Mandushev, Spassiva \& Staneva
(1991). A more direct method of determining the mass is to measure the
velocity dispersion of the stars in the clusters. Observations have
been obtained for nine clusters so far (two in NGC 1705 and one in NGC
1569 by Ho and Fillipenko 1996; two in M82 by Smith and Gallagher
2000, 4 in NGC 4038/39 by Mengel \etal\/ 2000). The dispersions range
from 10 - 20 km s$^{-1}$ and the masses range from 1 $\times$ 10$^5$
to 4 $\times$ 10$^6$ \msun\/, in good agreement with values for 
the more massive old globular clusters in the Milky Way.  The size and
dispersion measurements also show that the crossing times for the
clusters are $\approx$1 Myr. Hence, even the younger clusters have
survived many crossing times. The older clusters in NGC 7252, NGC
3921, NGC 4038/4039, and NGC 1275 ($\approx$ 500 Myr; see Tables 1 and
2), have survived for several hundred crossing times and appear to be
quite stable.  Their densities are $\approx$ 10$^5$ \msun\/ pc$^{-3}$,
similar to old globular clusters, hence these clusters will almost
certainly last for tens of Gyr.

\subsubsection{The Luminosity Function}

To first order, the luminosity functions of young compact clusters in
merging galaxies are power laws with index $\approx $ -2 (Table 1).
Harris and Pudritz (1994) have pointed out that
the mass function for giant molecular clouds is also a power law with
a similar index. Hence, all that may be necessary is a triggering
mechanism to get the molecular clouds to collapse and form star
clusters.  Jog \& Solomon (1992) have suggested that merger induced
starbursts can raise the ambient pressure in the ISM and trigger the
implosion of the molecular clouds. Elmegreen \& Efremov (1997) agree
that high-pressure environments are needed to trigger the star
formation and suggest other mechanisms might be high background virial
density (e.g., in dwarf galaxies), turbulent compression, or
large-scale shocks (in interacting galaxies).
 
The power law index for the young clusters is markedly different than
the Gaussian profile found for old globular clusters (e.g., Figure 3 of Zhang and Fall 1999,). However, various
destruction mechanisms (e.g., 2-body evaporation, bulge and disk
shocking, dynamical friction, stellar mass loss) should modify the
distribution with time. Two-body evaporation appears to be the
strongest amongst these mechanisms, destroying the fainter more
diffuse clusters first, and in certain conditions leaving a peaked
distribution similar to what is seen for old globular clusters (e.g.,
Fall and Zhang, 2000). This is similar to  young
clusters in the Milky Way with the OB associations typically only
lasting tens of Myr, and open clusters lasting hundreds of Myr.  Other
examples of clusters which are apparently dissolving are the Arches
and Quintuplet clusters near the Galactic Center, since no older
clusters are seen in their vicinity, and the $\approx$ 40 clusters in
the inner 6$''$ of NGC 7252, which all have ages less than about 10 Myr
(Miller \etal\/ 1997). Finally, the number of young clusters in the
Antennae galaxies is so large that it requires most of the clusters to
dissolve or the value of S$_N$ will be too high when it settles down
to become an elliptical (Whitmore \etal\/ 1999)!

\begin{figure*}
\includegraphics{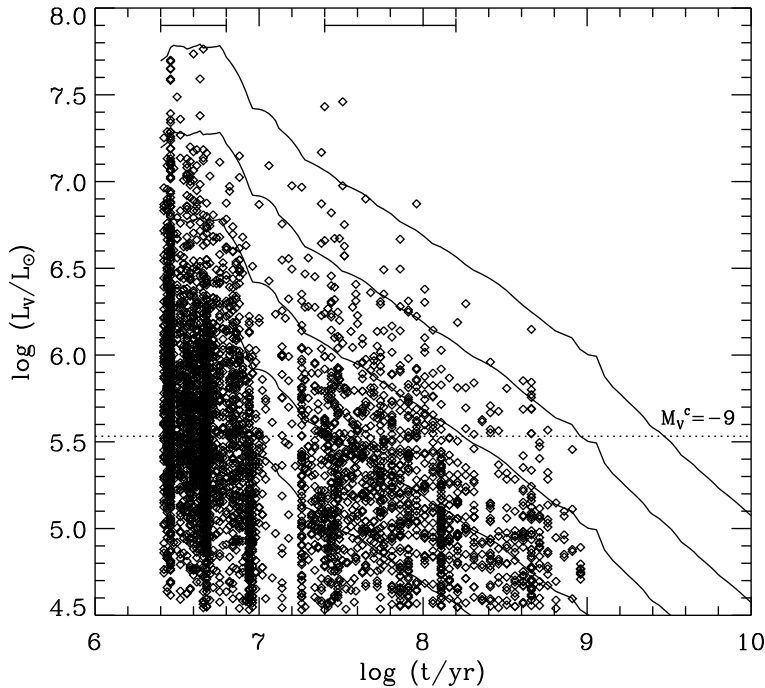}
\vspace{10cm}
\caption{Luminosity of cluster candidates in the Antennae as a 
function of their ages (from
Zhang and Fall, 1999,  Fig. 2). The lines represent the Bruzual-Charlot (1996) 
tracks with log(M/M$_{\odot}$) = 6.0 (top), 5.5, 5.0, 4.5, and 4.0. 
See Zhang and Fall for further details.
}
\end{figure*}

Fritze-v.Alvensleben (1999), following a similar line of reasoning to
Meurer (1995), has attempted to determine the mass function for the
clusters in the Antennae using the color information from the WFPC1
observations by Whitmore \& Schweizer (1995). They conclude that the
original mass function is a Gaussian which gets spread out in time to
form the power law luminosity function we observe today. However,
their analysis does not take into account the fact that the cutoff in
the observed luminosity function is due to incompletion at the faint
end (see Zhang \& Fall 1999 for a discussion). When convolved with
uncertain age estimate  used to convert from luminosity to
mass (e.g., due to reddening from dust and the
availability of only V-I colors), the resulting distribution will artificially appear to be
roughly Gaussian. A more complete treatment by Zhang \& Fall (1999),
using UBVI colors based on WFPC2 observations by Whitmore \etal\/
(1999) and corrections for reddening and incompletion, concludes that
the mass function is roughly a power law.

There is some evidence that the luminosity function for the young
clusters is not a perfect power law, but is steeper for bright
magnitudes (NGC 4038/4039 -
Whitmore \etal\/ 1999, NGC 3256 - Zepf \etal\/ 1999).  In the Antennae, Whitmore \etal\/ (1999) find that the
cluster luminosity function  appears to have a bend at $M_V \approx -10.4$ ($\approx$
--11.4 after making a correction for extinction).  For absolute
magnitudes brighter than $M_V \approx -10.4$ the power law is steep
and has an exponent of $\alpha = -2.6 \pm 0.2$, while for the range
$-10.4 < M_V < -8.0$ the power law is flatter, with $\alpha = -1.7 \pm
0.2$.  Assuming a typical age of 10 Myr for the clusters, and 1 mag of
extinction, the apparent bend in the LF corresponds to a mass $\approx
1 \times 10^{5}$ M$_{\odot}$, only slightly lower than the
characteristic mass of globular clusters in the Milky Way. A similar bend
may be present in the mass function derived by Zhang \& Fall (1999). The bend may be a precursor to
what will become the peak of the globular cluster luminosity function.

\subsection{Young Compact Star Clusters in Starburst Galaxies}

Young compact star clusters are also found in many starburst galaxies,
but in much smaller numbers than the merging galaxies.  Meurer et
al. (1992), using ground-based observations, found a population of
young clusters in the nearby starburst dwarf galaxy NGC 1705, the
brightest of which was an unresolved off-center nucleus which they
proposed as a young (13 Myr) globular-like cluster with a mass 
$\approx$ 1.5 $\times$ 10$^6$ \msun. Meurer \etal\/ (1995) followed
this up with an extensive study of nine starburst galaxies obtained
with the Faint Object Camera on HST. All nine of the galaxies contained young
compact star clusters. On average, 20 \% of the UV light from the
galaxies comes from the clusters. The brightest clusters are preferentially found near the centers of the
galaxies. They find the sizes are similar to Galactic globular
clusters and the luminosity function has an index $\approx$ -2. Hence,
the clusters found in the starburst galaxies appear to be similar to
the clusters found in merging galaxies.

Several other authors find similar examples in other starburst
galaxies, as listed in Table 3. Conti \& Vacca (1994) observered 19
``knots'' in the Wolf-Rayet galaxy He 2-10, each with a luminosity,
mass, and size similar to Galactic globular clusters. Other early
observations include those of O'Connell \etal\/ (1994) for NGC 1569 and
NGC 1705, Hunter \etal\/ (1994) for NGC 1140, and Watson \etal\/ (1996)
for NGC 253.

The case of M82, the prototypical starburst dwarf galaxy, is especially
interesting. O'Connell \etal\/ (1995) find a complex of over 100
compact, luminous ``super star clusters'' concentrated in the inner
100 pc of the galaxy shining though a relatively dust free region.
The brightest cluster has M$_V$ = -13.2 while the mean M$_V$ is -11.6
mag. Since most of this galaxy is embedded in dust the total number of
young clusters is likely to be several times this value. It is quite
possible that the starburst in M 82 was triggered by a tidal
interaction with its larger neighbor, M 81, hence it is not clear
whether M 82 (or several other starburst galaxies with evidence for
interactions) should be in this section or in the previous section on
interacting galaxies. De Gris, O'Connell, \& Gallagher (1999) 
studied a region farther from the center of M82 where active star
formation is not occuring.  They estimate the ages of the clusters in
this region at 20 - 100 Myr. They find that
the objects in the outer regions have sizes in the range 2.3 $<$ \reff\/ $<$ 8.4
pc. While the lower value is similar to the sizes of galactic globular
clusters, the higher value is more typical of open clusters. They also find that the brightest clusters have
M$_V$ $\approx$ -10 mag, and most are in the range -5 to -7 mag. Hence,
most of these clusters are too faint to become globular clusters since
they will fade several magnitudes as the stars evolve. It appears that
this region is not able to form the true ``super star clusters'' seen
near the center of M82 and in other merger and starburst systems.

The lesson appears to be that luminous
young star clusters are found whenever there is vigorous star
formation, whether it be in mergers or starburst galaxies.  Since the ultraluminous IRAS sources are essentially all
mergers (Sanders \etal\/ 1988), it is not surprising that mergers show
the largest populations of young star clusters.

\subsubsection{Young Compact Star Clusters in Barred Galaxies}

Barth \etal\/ (1995) found young  clusters in the circumnuclear star-forming rings around the barred spiral galaxies NGC 1097 and NGC 6951. The clusters are compact, with \reff\/ $\approx$ 2.5 pc in NGC 1097 and $\leq$ 4 pc in NGC 6951. The brightest cluster has M$_V$(uncorrected for extinction) = -12.6. They estimate an intrinsic
M$_V$ in the range -14 to -15, since the clusters are on the outer edges of prominent dust lanes. Hence, these clusters appear to be quite similar to 
clusters in merging and starbursting galaxies.
Maoz \etal\/ (1996) found a similar population of clusters in a sample of five barred 
galaxies using the Faint Object Camera. They estimate that 10 - 40 \% of the UV light is
from compact clusters in these galaxies, similar to the result for
starburst galaxies (Meurer \etal\/ 1995).
Buta \etal\/ (2000) have done a careful analysis of young compact
clusters in the nuclear ring of NGC 1326, an early-type barred spiral
in the Fornax cluster.  They find 269 candidate clusters with ages in
the range 5 to 200 Myr, but no  clusters older than
this. The older clusters still lie within the ring, with no evidence
of migration. The luminosity function has an index of -2.1, similar to
the other compact clusters discussed in this review, but the brightest
clusters are fainter than the brightest clusters found in
mergers or starburst galaxies, with no M$_{V}$ (uncorrected) brighter than -11. The authors conclude that this galaxy lacks any true
super star clusters, and suggest that super star clusters are not a
universal property of star-forming rings. It is  interesting to note that while the
strong Lindblad resonance in this galaxy can produce clusters with a
typical power-law luminosity function, it apparently cannot form the
brightest clusters which are the best candidates for protoglobular
clusters.

\subsection{Young Star Clusters in Spiral Galaxies}

Larsen \& Richtler (1999) carried out a systematic ground-based search
for young massive clusters in 21 nearby non-interacting 
spiral galaxies and found young massive star clusters in about one
quarter of the galaxies.  In a followup paper (Larsen \& Richtler
2000), they add a variety of other galaxies to the sample from the literature, including
merging and starbursting galaxies, in order to test what conditions are
most advantageous for making large numbers of massive clusters. They
define the {\it specific cluster frequency} (not to be confused with the specific globular cluster frequency, S$_N$, see Harris 1991) as the fraction of light in clusters to the fraction of light in
the total galaxy: T$_L$ = 100 $\times$ L$_{clusters}$ / L$_{galaxy}$.

They prefer to make the measurement in U which is most sensitive to
young clusters. Their primary result is that T$_L$(U) is well
correlated with the star formation rate per unit area (Figure
8). Galaxies with very active star formation form proportionally
more of their stars in clusters  than in the field, with some
merger and starburst galaxies devoting as much as 15 - 20 \% of their
luminosity to clusters.  Note that this is precisely what is needed to
increase the specific globular cluster frequency, a concern voiced by Harris
(1999).  Larsen \& Richtler (2000) also argue that ``The cluster
formation efficiency seems to depend on the SFR in a continuous way,
rather than being related to any particularly violent mode of star
formation''. 

Closer to home, Chandar \etal\/ (1999) have used WFPC2 observations to
study the young compact clusters in M33. They finds 44 young clusters
with ages $\leq$ 100 Myr and masses in the range 6 $\times$ 10$^2$ to
2 $\times$ 10$^4$ \msun\/. Hence, M33 appears to currently be making many
young compact clusters, but few if any with the masses of regular
globular clusters.

\begin{figure*}
\includegraphics{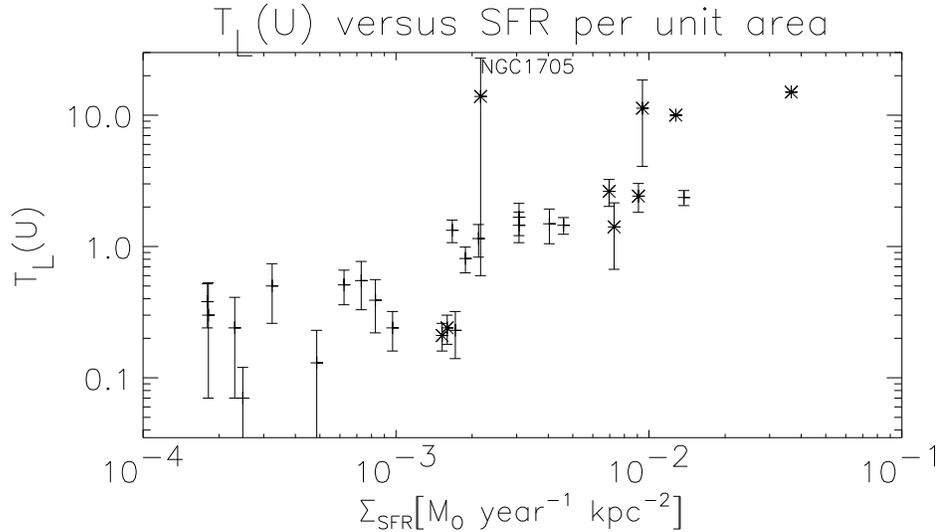}
\vspace{8cm}
\caption{Plot of the specific luminosity for the clusters in the U
band  vs. the star formation rate per unit area
for a sample of spiral, starburst, and merging galaxies (from Larsen \& Richtler \etal\/ 2000, Fig 6).}
\end{figure*}

\subsubsection{Young Compact Star Clusters in Tidal Tails}

Knierman \etal\/ (2000) have examined six tidal tails in four 
prototypical mergers (NGC 3256, NGC 3921, NGC 4038/4039 and NGC
7252). They find that only one of the tails (the western tail of NGC
3256) currently has a large number of young compact clusters (i.e.
$\approx$ 50 clusters with the brightest having M$_V$ $\approx$ -10
mag). It is not clear whether the clusters were formed when the tidal
tail was ripped from the galaxy or are currently forming. Some of the other tails appear to have
only a few young clusters, (e.g., NGC 7252 and NGC 3921) while others
(e.g., NGC 4038/39) appear to have essentially no clusters in the
tails.  Hence, it appears that there is a wide range in the number of
clusters in tidal tails, perhaps due to differences in how the tails
were generated (e.g., gas-rich versus gas poor, deep penetrating orbit
versus quiescently being pulled out from the outer regions of the
galaxy, etc.). Other studies including observations
of young compact star clusters in tidal tails include Lee, Kim \&
Geisler (1997), Tyson \etal\/ (1998), and Gallagher \etal\/ (2000).

% Polar rings

\section{A Compilation of the Literature and a Discussion of Broader Issues}

Tables 2, 3, and 4 provide a compilation of HST (and occasionally key
ground-based) observations of young compact star clusters in merging,
starbursting, and miscellaneous other galaxies.

Based on the discussion in \S 2 and 3, and the compilation in Tables
1, 2, 3, and 4, it is clear that luminous young compact star clusters
are produced in a wide variety of environments, but in much greater
number in mergers and starburst galaxies, systems where vigorous star
formation is occurring. A similar conclusion was reached by van den
Bergh (2000), who comments, ``Presently available data strongly suggest
that the specific cluster forming frequency is highest during violent
bursts of star formation''. In addition, the most luminous clusters
are formed in the regions with the most violent star formation.

An important question is whether this is a statistical effect due to
the lower number of clusters in galaxies with low star formation, or
whether it is physically more difficult to form massive clusters in
relatively quiescent systems (i.e., ``is there a cutoff at the high end
of the luminosity function for quiescent galaxies? ''). The situation may
be analogous to the upper IMF in 30 Doradus. It was presumed that the
large number of very luminous stars indicated that conditions in 30
Doradus were especially conducive for making high mass stars. However,
Massey \& Hunter (1998) find that the IMF is normal; that the large
number of massive stars is simply due to the tremendous number of
stars in the system and the young age of the cluster.

The most straightforward approach to answering this question would be
to look at the mass function of the clusters for a variety of
galaxies. Unfortunately, this is quite difficult given the large
amounts of dust and the dimming caused by stellar evolution. The only
galaxy where this has been attempted in detail is the Antennae (Zhang
\& Fall 1999). However, we can attempt to make the comparison using
the luminosity function, as shown in Figure 9 for 8 galaxies. We
find that all  the galaxies have luminosity functions with similar
slopes, with an average power law index $\alpha$ = -1.93 $\pm$ 0.06 (uncertainty in the mean; the scatter is 0.18). The primary
difference is the normalization of the luminosity function, with NGC 3256 and NGC 4038/39 having large numbers of clusters while NGC 3921 and HE 2-10
have relatively few clusters. There is
no obvious trend for a cutoff at high luminosity for the more
quiescent galaxies, suggesting a universal luminosity function is a reasonable
approximation.

Such an approach is oversimplified for a number of reasons, primary
amongst them being that the luminosity of the clusters vary with
time. For example, a single-age burst population will evolve to the right in
Figure 9, making it difficult to determine whether the luminosity
function is lower because of evolution or due to a smaller number of
clusters originally forming. Other difficulties with this simple model are that
it assumes similar star formation histories for the various galaxies
(e.g., continuous rather than sporadic bursts at
different times), and ignores the fact that the faint end will probably
undergo rapid evolution as the faint clusters dissolve.  Nevertheless,
to first order the luminosity functions appear to be remarkably
similar in form.

\begin{figure*}
\includegraphics{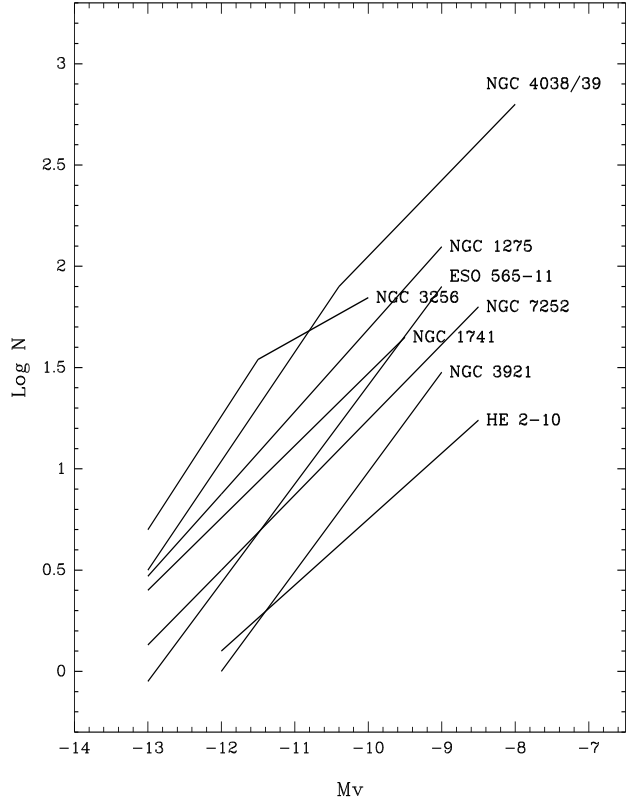}
\vspace{11cm}
\caption{Approximate luminosity functions for galaxies in Table 1, normalized to have 0.25 mag bins.
}
\end{figure*}

\begin{table}
\begin{center}
\caption{Observations of Interacting Galaxies with Young Star Clusters}
% \begin{tabular}{lll}
\begin{tabular}{p{1.7in}p{3.2in}}
\hline 
 Reference  & Brief Description \\
\hline

% \multicolumn{2}{c}{(Mergers)} \\

Schweizer (1982) & NGC 7252 (ground-based, 6 knots, stat. significant?)\\
Lutz (1991) & NGC 3597 (ground-based, $\approx$ 10 knots, lacked resolution) \\
Holtzman et al. (1992) & NGC 1275  (proposed ``protoglobular clusters'', n = 60)\\
Whitmore et al.  (1993) & NGC 7252 (prototypical merger, n = 40)\\
Crabtree et al. (1994) & NGC 7727 (ground-based) \\
Whitmore \& Schweizer (1995) & NGC 4038/4039 (n = 700, Antennae galaxies)\\
Zepf et al. (1995) & NGC 1275 (ground-based spectra, .1 - 1 Gyr)\\
Borne et al. (1996) & Cartwheel galaxy (clusters in rings) \\ 
Holtzman et al.  (1996) & NGC 3597, NGC 6052  (mergers, not cooling flows)\\
Schweizer et al. (1996) & NGC 3921  (102 candidate globulars, 49 ``associations'') \\
Hilker \& Kissler-Patig (1996) & NGC 5018 (several hundred Myr to 6 Gyr)  \\
Miller et al. (1997) & NGC 7252 (n = 499, 3 pop., $<$ 10 Myr for R $<$ 6$''$) \\
Whitmore et al.  (1997) & NGC 1700, NGC 3610 (missing link with ellipticals ?)\\
Schweizer \& Seitzer (1998) & NGC 7252 (spectra, n = 8, ages, metallicities)\\
Brodie et al. (1998) & NGC 1275 (ground-based spectra, age $\approx$ 450 Myr)\\
Carlson et al. (1998) & NGC 1275 (n = 3000, mix of red and blue clusters)\\
Johnson et al. (1998) & NGC 1741 (starburst, interacting, Hickson group)\\
Stiavelli et al. (1998) & NGC 454 (5-10 Myr, effects of emission on photometry)\\
Dinshaw et al. (1999) & NGC 6090 (n = 4, NICMOS observations)\\
Zepf et al. (1999) & NGC 3256 (n = 1000, 15-20 \% of U light, break in LF ?)\\
Whitmore et al.  (1999) & NGC 4038/4039 (n=800 to 8000, break in LF ?)\\
Gallagher et al. (2000a) & Stephan's Quintet (n = 150, galaxies and tidal tails)\\
Alonso-Herrero et al. (2000) & Arp 299 (ULIRG, n = 40)\\
Forbes \& Hau (2000) & NGC 3597 (ground-based, K band, $\alpha$ = -2)\\
Johnson \& Conti (2000) & HCG 31 (several in Hickson Compact Group 31)\\
Gilbert et al. (2000) & NGC 4038/39 (IR spectra, ages, masses)\\
Mengel et al. (2000) & NGC 4038/39 (IR spectra, ages, masses)\\ 
Georgakakis et al. (2000) & NGC 6702 (dust-lane elliptical, 2- 5 Gyr)\\
Goudfrooij et al. (2000) & NGC 1316 (elliptical with shells, 3 Gyr)\\

\end{tabular}
\end{center}
\end{table}

\begin{table}
\begin{center}
\caption{Observations of Starburst Galaxies with Young Star Clusters}
% \begin{tabular}{lll}
\begin{tabular}{p{1.5in}p{3.4in}}
\hline 
 Reference  & Brief Description \\
\hline

Arp \& Sandage (1985) & NGC 1569 (ground-based, coined ``super star clusters'') \\
Kennicutt \& Chu (1988) & LMC (cores of HII regions may be globular clusters) \\
Meurer et al. (1992) & NGC 1705 (ground-based, 10$^6$ \msun)  \\
Conti \& Vaca (1994) & He~2-10 (Wolf-Rayet galaxy, 1 - 10 Myr, 10$^5$ - 10$^6$ \msun)\\
Hunter et al. (1994) &  NGC 1140  (n = 7, merger ?, 3 - 15 Myr) \\
O'Connell et al. (1994) & NGC 1569, 1705 (n=3, \reff $\approx$ 3 pc, density $>>$ R136) \\
Meurer et al. (1995) & 9 starbursts (20 \%  of UV from clusters, $\alpha$ = -2)\\
O'Connell et al. (1995) & M82 (n $\approx$ 100, \reff\/ = 3.5 pc, near center) \\
Watson et al. (1996) & NGC 253 (n = 4, brightest = -15 mag and 1.5 $\times$ 10$^6$ \msun\/) \\
Leitherer et al. (1996) & NGC 4214 (FOC and FOS, n = 200, 4-5 Myr)\\
de Marchi et al. (1997) & NGC 1569 (1569A is superposition of two clusters) \\
Ho \& Fillppenko (1997) & NGC 1705, NGC 1569 (spectra, velocity
disp., 3.3 $\times$ 10$^5$ \msun\/) \\
% Gorjian (1997) & NGC 5253 ( \\
Calzetti et al.(1997) & NGC 5253 (BCG, n = 6, 2.5 Myr, $\approx$ 10$^6$ \msun\/) \\
Oslin et al. (1998) & ESO338-IG04 (BCG)\\
De Grijs et al. (1999) & M82 (outer region, 20 - 100 Myr, fainter than -10 mag)\\
Gallagher et al. (2000) & NGC 7673, NGC 3310, Haro I (clumps of SSCs)\\
Johnson et al. (2000) & He~2-10 (Wolf-Rayet galaxy, WFPC2, H$_{\alpha}$, GHRS)\\
Smith et al. (2000) & M82 (ground-based spectra, 60 Myr, 2 $\times$ 10$^6$ \msun\/) \\
Oslin (2000)  & Mrk 930, ESO185-IG13, ESO350-IG38 (BCGs)\\
Meurer (2000) & NGC 3310 (0 - few 100 Myr, continuous formation)

\end{tabular}
\end{center}
\end{table}

\begin{table}
\begin{center}
\caption{Observations of Other Galaxies with Young Star Clusters}
% \begin{tabular}{lll}
\begin{tabular}{p{1.5in}p{3.4in}}
\hline 
 Reference  & Brief Description \\
\hline

Barth et al. (1995) & NGC 1097, NGC 6951 (barred, 2 - 3 pc, n = 88 and 24) \\
Holtzman (1996) & Abell 496, 1795, 2029, 2597 
(not related to cooling flows)\\
Maoz et al. (1996) & 5 barred (FOC observations, 10 - 40 \% of UV from clusters) \\ 
Lee et al. (1997) & UGC 7636 (dwarf near NGC 4472, n = 18, tidal tail) \\
Carollo et al. (1997) & 35 spirals (M$_V$ vs. R$_e$ diagram)\\
Tyson et al. (1998) & NGC 5548 (in tidal tail of Seyfert galaxy)\\
Buta et al. (1999) & ESO 565-11 (barred, n = 700, 4-6 Myr, $\alpha$ = -2.2)\\
Chandar et al. (1999) & M33 (n=44, 10$^3$ - 10$^4$ \msun\/, $<$ 100 Myr \\ 
Buta et al. (2000) & NGC 1326 (barred, n = 269, 5 - 200 Myr)\\
Larsen \& Richtler (2000) & 21 spirals (ground-based spirals, also mergers \& starbursts) \\
Gallagher et al. (2000) & Stephans' Quintet (Hickson compact group)\\
Knierman et al. (2000) & Tidal tails in 4 mergers (not all have clusters)\\

\end{tabular}
\end{center}
\end{table}

\begin{figure*}
\includegraphics{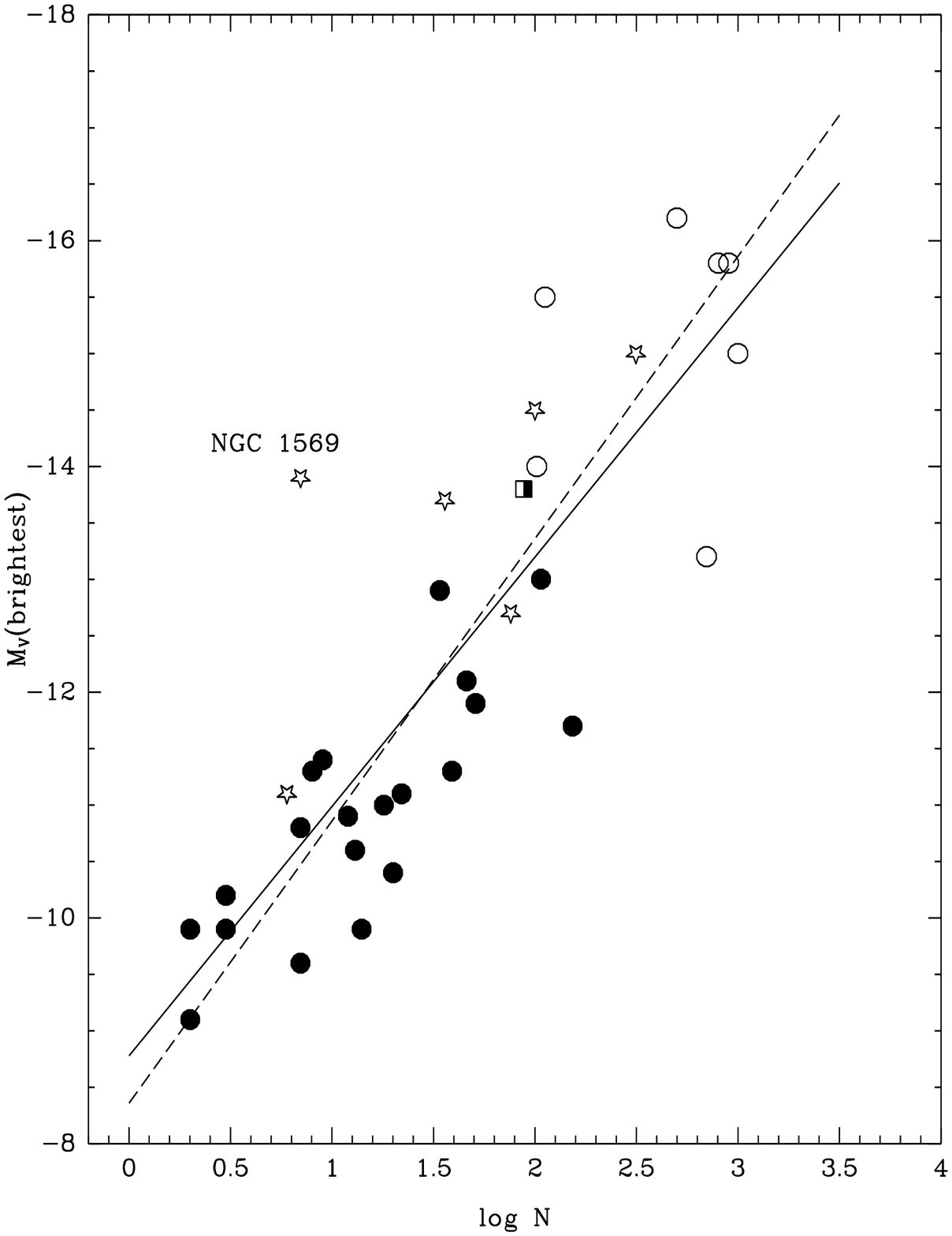}
\vspace{11cm}
\caption{Plot of the magnitude of the brightest cluster vs. 
the log of the number of clusters. Filled circles are spiral galaxies
from Larsen \& Richtler (2000), open circles are mergers, stars are starbursting
galaxies, and the half filled square is a barred galaxy (Table 1). The solid
line is a best fit (excluding NGC 1569) while the dashed line is the prediction
from a universal power law luminosity function with index $\alpha$ = 2.
See text for more details.
}
\end{figure*}

Another approach which allows us to increase the sample at the expense of more scatter for any particular galaxy is to plot the
magnitude of the brightest cluster vs. the number of clusters in the
galaxy, as shown in Figure 10.  This figure uses the groundbreaking
survey of Larsen \& Richtler (2000), with the additions of some new
points for merging and starburst galaxies drawn from the papers in
Tables 2, 3, and 4.  We find a clear trend between the number of
clusters observed and the magnitude of the brightest cluster. The
solid line is the fit to the data (excluding NGC 1569) with a slope =
-2.3 $\pm$ 0.2 . The dotted line shows the trend expected if there is a
universal luminosity function with $\alpha$ = -2 and the increase in
the luminosity is simply due to a larger sample of clusters (i.e., the
slope is -2.5). Again, to first order it appears that a universal luminosity function can
explain the data, even with the large scatter expected from low number
statistics, non-uniform databases, differences in selection criteria,
and differences in cutoff magnitudes (only those with cutoffs
$\approx$ -9 have been included).

However, this may not be the whole picture. It is easy to think of
examples that do not appear to fit this picture. For example, the
brightest clusters in NGC 1569 and NGC 1705 are 2 - 3 magnitudes
brighter than the second brightest cluster in the galaxy (O'Connell, Gallagher \& Hunter
1994, see Figure 9 of Meurer \etal\/ 1995 which provides a graphical
representation for NGC 1705), suggesting something special is
happening in these clusters. In addition, many galaxies are currently
forming  a few very young compact clusters (e.g, the central regions of the Milky Way and NGC 7252) which, assuming a steady formation rate over a long period of time, implies that a few very massive clusters
should eventually form, based on the statistics. These appear to be
missing. However, it is possible that the young clusters are forming in regions that are not conducive to the long-term survival of the clusters.

\section{Current and Future Questions}

The HST observations of star clusters have answered many questions,
but typical of any active field of science, they have introduced
even more new questions.  Here are some of the fundamental
questions that should be addressed over the coming decade.

\subsection{Will some of the young compact clusters survive to become
classical globular clusters, and if so, how many ?}

Historically, astronomers have been approaching this question from two
different directions. Looking at resolved clusters, Kennicutt and Chu
(1988) concluded ``that populous clusters may be forming in giant HII
regions, but only a small fraction of giant HII regions are likely to
contains such clusters''. The prototype for this idea is 30
Doradus. For the more distant galaxies where the clusters are not well
resolved, most of the original motivation came from trying to
understand the specific frequency of globular clusters in elliptical
galaxies (Schweizer 1982, Burstein, 1982, Ashman and Zepf 1992).  As
pointed out in \S 3.1.5, we actually need the vast majority of
clusters to dissolve or we end up with specific globular cluster
frequencies that are too high.  

It now seems well established that some of the young compact clusters
will survive to form globular clusters. For example, in NGC 7252 and
NGC 3921 the clusters are already 500 Myr (several hundred crossing
times), have the distribution expected of globular clusters, and have
the same masses and densities of classical globular clusters.  The
remaining question is {\it how many} of the clusters will survive and
become old globular clusters. In particular, is this how the red
(metal-rich) population of globular clusters found in elliptical
galaxies are formed, as Ashman \& Zepf (1992) propose, or is this just
a minor trace population? For example, Schweizer \etal\/ (1996)
concludes that the total number of globular clusters in NGC 3921 has
increased by only 40 \%. While sizeable, this may not be enough to
explain the increase in \sn\/ in ellipticals unless the typical
ellipticals has several major mergers in its lifetime.  In addition, a
clear prediction from Ashman \& Zepf is that the ratio of red
(metal-rich) to blue clusters should increase for high \sn\/ galaxies.
However, Forbes, Brodie, \& Grillmair (1997) find that the number of
red clusters does not increase with \sn.  It is possible that this is
because their sample is dominated by ellipticals in clusters of
galaxies, where other mechanisms might also be operating (e.g.,
stripping the globular clusters out of nearby dwarf galaxies; see the
Harris review in this volume). In summary, it appears that many of the
brighter young compact clusters will become classical globular
clusters, but the jury is still out on whether this is the cause of
the increase in \sn\/ for elliptical galaxies.

\subsection{What fraction of stars are formed in clusters?}

Since only a subsample of the young clusters are likely to survive, an
obvious question is whether most of the field stars in a galaxy are
originally formed in clusters. In the Milky Way, approximately 0.1\% of
the stars are currently in globular clusters.  However, in some
starbursting and merging galaxies the fraction of light from the 
clusters is as high as 20 \% (Meurer \etal\/ 1995).  Even in these
young star forming regions many of the field stars are from clusters
that have already dissolved, hence the true percentage of stars that
were originally in clusters is even higher, and might conceivably
be $\approx$ 100 \%.

HST observations may allow us to answer this question by determining
the rate at which clusters dissolve. For
example, if we were to assume that the Antennae has been making
clusters at the same rate for the past 200 Myr (a rather uncertain
assumption to say the least), we could use figure 2 from Zhang and
Fall (reproduced as Figure 7) to show that for every 20 clusters
originally formed, only about 1 will survive to an age of $\approx$ 100
Myr (i.e. there are roughly the same number of clusters
in the 0 - 10 Myr age bin as in the 20 - 200 Myr age bin). 
While this very crude calculation is probably not justifiable for
a single galaxy, which we may be catching during the peak of cluster
formation, once a larger sample becomes available this would be a
reasonable approach. 

An intriguing result is the finding that the luminosity
function in the Ursa Major dwarf spheroidal galaxy and the globular
cluster M15 are essentially indistinguishable (Wyse \etal\/ 1999), even
though the densities differ by three order of magnitude. It is
tempting to suggest that perhaps most the stars are formed in groups
and clusters, and that the field stars are simply the remnants of the
fainter, less dense clusters which have dissolved.

\subsection{What fraction of star formation is triggered by other star formation?}

There appears to be a variety of ways to form stars (e.g.,
gravitational instabilities, shocks between colliding clouds of gas,
enhanced pressure of the ISM, etc.). As discussed in \S 2.3, HST
observations suggest that star formation can also be triggered by
nearby bursts of star formation (e.g., around 30 Doradus; see Figures
1 and 2). An interesting question is what fraction of all stars have
been formed this way? At any one time 
only a relatively small fraction of star formation appears to be triggered (e.g.,
the clusters around 30 Doradus are relatively modest compared to 30
Doradus itself). However, in principle this is a self-propagating
process which may continue over a much longer period of time, hence it
is possible that overall a relatively large fraction of star formation is
triggered. The fact that much of the triggered star formation is still
embedded in dust clouds makes it difficult to obtain a
complete census. New observations with the NICMOS + cryocooler, and the IR
channel of the WFC3, will help answer the question of how important this
mechanism is to the total production of stars in a galaxy.

\subsection{Is a massive open cluster
the same as a low-mass globular cluster ?}

The fact that the luminosity functions for young clusters are power laws begs the question of whether there is anything
fundamentally different between a massive open cluster and a low-mass globular
cluster.  Are we looking at a continuum,  or a bimodal distribution
with fundamentally different formation mechanism for open clusters and
globular clusters?  It seems possible that the distinction between the
two is artificial, and is due to the fact that we live in a galaxy
which had an initial burst of star formation 14 Gyr ago but no major
bursts since then.
The only clusters that have  survive from
the initial burst are, by necessity, massive and
compact. These we call globular clusters. 
In the present epoch, the star formation rate is percolating
at a much lower rate and we are only able to see the spectrum of
clusters from associations to open cluster. We do not see  young
globular clusters for several possible reasons. First, they should
only form very rarely, since the star formation rate is so low (see \S
4). Second, we would probably call them open clusters anyway, since
we are not use to calling anything young a globular cluster. Indeed,
there is overlap in the masses of open and globular clusters.
Candidate open cluster/globular clusters might include
M67 (5 Gyr), Be 17 and
Lynga 7, which according to Phelps \etal\/ (1994), may be
as old as the youngest globular clusters.

\subsection{Can we develop a unified picture of cluster formation that
explains all this?}

While we are making good progress understanding many pieces of the
puzzle, how it all fits together is a much tougher question.  Is it
possible to develop a universal model that provides a framework for
understanding cluster formation both near (e.g., the classic picture
of associations, open clusters, and globular clusters in the Milky
Way), intermediate (``populous'' star clusters in the LMC and ``super
star clusters'' in nearby dwarf starbursts), and far (the young
compact clusters in mergers and starbursts); for spiral and elliptical
galaxies; for the initial collapse of a many galaxies $\approx$ 14 Gyr ago
and mergers of galaxies in the local universe; and for violent
starbursts and ``quiescent'' star formation?  Some of the ideas
discussed in this review lead to the following working hypothesis,
portions of which several groups are pursuing, in particular Elmegreen
\& Efremov (1997), Vesperini (1998), and Fall \& Zhang (2000).

The luminosity functions for young clusters (e.g., Whitmore \&
Schweizer 1995), molecular clouds (Harris \& Pudritz 1994) and HII
regions (Elmegreen \& Efremov 1997) are all power laws with index
$\approx$ -2 . Hence, we start with a universal power law luminosity
(mass) function for. The total number of clusters is normalized
depending on how active the star formation is (e.g., Larsen \&
Richtler, \S 4). This explains the existence of large numbers of
bright clusters in mergers, since they have the most active star
formation.  The physics of how the cluster formation is triggered is
still uncertain, but several possible mechanisms have been proposed(e.g., Jog \&
Solomon 1992, Kumai \etal\/ 1993, Elmergreen \& Efremov 1997). The
power law evolves due to both internal (e.g., evaporation, stellar mass loss)
and external (e.g., tidal stress, triggered star formation) influences
(e.g., Fall \& Zhang 2000), with most of the faint and/or diffuse
clusters dissolving, just as they do in the Milky way. This model would then need to be
convolved with models of galactic evolution (i.e., a
combination of initial collapse, hierarchial merging, and
internal galactic dynamics), and stellar evolution (dimming and
reddening of the starlight) to produce the wide variety of cluster
demographics we see in galaxies.

\subsection{What is the limiting redshift for observing young globular clusters
with HST and NGST?}

The current limiting redshift for observing young globular clusters
with characteristics similar to the brightest young clusters in the Antennae is Z $\approx$ 0.5, using the WFPC2 on HST.  It will be
possible to do slightly better (i.e.  Z $\approx$ 0.8), with the
Advanced Camera for Surveys since it has a quantum efficiency which
is roughly 3 times better than WFPC2. However, the real
breakthrough will be NGST, where Burgarella \& Chapelon (1998)
estimate that it will be possible to observe young globular clusters
out to Z $\approx$ 9, if they exist. Since globular clusters appear to
be the oldest fossils we observe in galaxies it is quite possible
that the first objects we will see emerging from the ``dark ages''
will be young compact star clusters, similar to what we are seeing in
nearby galaxies.

\end{document}